\documentclass[journal]{IEEEtran}
\usepackage{stackengine}
\usepackage{amssymb}
\usepackage{amsmath}
\allowdisplaybreaks
\usepackage{dsfont}

\newcommand \obar[1]{\stackon[1.1pt]{$#1$}{\rule{1.2ex}{0.075ex}}}
\usepackage{amssymb}
\usepackage{cite}
\usepackage{graphicx}
\usepackage{bm}
\usepackage{multirow}
\usepackage{booktabs}
\usepackage{color}
\usepackage{upgreek}
\usepackage[bookmarks={false}]{hyperref}
\usepackage[hyphenbreaks]{breakurl}
\usepackage{balance}
\usepackage{verbatim}
\usepackage{relsize}
\usepackage{nomencl}
\usepackage{ifthen}
\usepackage{soul}

\usepackage{accents}
\usepackage{setspace}
\usepackage{bbold}
\usepackage[compact]{titlesec}
\usepackage{tikz}
\usepackage{amssymb}
\usepackage{cuted}
\usepackage{float}
\usepackage{nicematrix}
\usepackage{subfigure}
\usepackage[linesnumbered,ruled,vlined]{algorithm2e}

\begin{document}

\title{DERs-Aided Blackstart and Load Restoration Framework for Distribution Systems Considering Synchronization and Frequency Security Constraints}

\author{Salish~Maharjan,~\IEEEmembership{Member,~IEEE,}        
        Cong~Bai,~\IEEEmembership{Student~Member,~IEEE,}
        Han~Wang,
        Yiyun~Yao,
        Fei~Ding,a~\IEEEmembership{Member,~IEEE,}
        and~Zhaoyu~Wang,~\IEEEmembership{Senior~Member,~IEEE}
\thanks{This work was funded by the U.S. Department of Energy Solar Energy Technologies Office under Agreement Number 40385}
\vspace{-1cm}}

\markboth{Journal of \LaTeX\ Class Files,~Vol.~14, No.~8, August~2015}%
{Shell \MakeLowercase{\textit{et al.}}: Bare Demo of IEEEtran.cls for IEEE Journals}

\maketitle

\begin{abstract}
Extreme weather events have led to long-duration outages in the distribution system (DS), necessitating novel approaches to blackstart and restore the system. Existing blackstart solutions utilize blackstart units to establish multiple microgrids, sequentially energize non-blackstart units, and restore loads. However, these approaches often result in isolated microgrids. In DER-aided blackstart, the continuous operation of these microgrids is uncertain due to the finite energy capacity of commonly used blackstart units, such as battery energy storage (BES)-based grid-forming inverters (GFMIs). To address this issue, this article proposes a holistic blackstart and restoration framework that incorporates synchronization between microgrids and the entire DS with the transmission grid (TG). To support synchronization, we leveraged virtual synchronous generator-based control for GFMIs to estimate their frequency response to load pick-up events using only initial/final quasi-steady-state points. Subsequently, a synchronization switching condition was developed to model synchronizing switches, aligning them seamlessly with a linearized branch flow problem. Finally, we designed a bottom-up blackstart and restoration framework that considers the switching structure of the DS, energizing/synchronizing switches, DERs with grid-following inverters, and BES-based GFMIs with frequency security constraints. The proposed framework is validated in IEEE-123-bus system, considering cases with two and four GFMIs under various TG recovery instants.
\end{abstract}


\IEEEpeerreviewmaketitle
\section{Introduction}
Transmission grids (TG) can be vulnerable to extreme weather events, which can result in blackouts in the downstream Distribution System (DS). Traditionally, distribution operators have relied on diesel generators (DGs) to blackstart and restore power during prolonged outages, ensuring a continuous power supply by maintaining a steady fuel source. With the anticipated widespread adoption of renewable energy sources in the DS, there is considerable interest in utilizing distributed energy resources (DERs) such as battery energy storage (BES) and renewables for blackstarting operations. Successfully implementing this technology would significantly reduce DS operators' reliance on DGs, which are known for their high operating costs.
\par
DS utilities subscribe to weather forecasting agencies to predict extreme weather events, which aids in pre-event preparation where blackstart and load restoration plans are essential \cite{APPA}. Blackstart planning with multiple BES inverters poses several challenges: (a) BES-based grid-forming inverters (GFMIs) must operate within frequency security constraints, (b) their finite energy capacity must be optimally utilized to establish cranking paths for activating renewable-based grid-following inverters (GFLIs) while supplying non-switchable loads, (c) optimizing synchronizing decisions to aggregate the energy and power capacity of BES-based GFMIs for rapid load restoration, and (d) ensuring synchronization with the TG upon availability to maintain continuous operation.
\par
Broadly, blackstart and load restoration strategies in the DS can be categorized into two main approaches, studied through: (a) determining the final configuration \cite{Li2014DistributionSearch,Sharma2015AIslanding,Wang2019CoordinatingSystems}, and (b) sequencing configurations \cite{Chen2018SequentialMicrogrids,Ding2022ANetworks,Liu2021CollaborativeDERS,Arif2022SwitchingRestoration,Chen2019TowardRestoration,Gao2022DecentralizedResources,Che2019AdaptiveConditions,Zhang2021AConstraints,Du2022Black-StartMicrogrids,Liu2023UtilizingRestoration}. The former works focus on establishing the final network topology without detailing the step-by-step evolution needed to achieve this configuration, which can complicate implementation. In contrast, the latter approach is more practical as it provides a sequence of feasible configurations to guide operators during the blackstart process. Furthermore, the latter works can be sub-classified into studies that either neglect \cite{Chen2018SequentialMicrogrids,Ding2022ANetworks,Liu2021CollaborativeDERS,Arif2022SwitchingRestoration,Chen2019TowardRestoration,Gao2022DecentralizedResources} or incorporate \cite{Che2019AdaptiveConditions,Zhang2021AConstraints,Du2022Black-StartMicrogrids,Liu2023UtilizingRestoration} frequency constraints in their restoration models.

\begin{figure}
    \centering
    \includegraphics[width=0.75\linewidth]{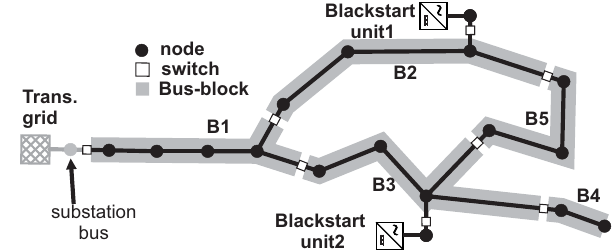}
    \caption{Example DS illustrating bus-blocks, switches, and blackstart units.}
    \label{fig_example_DN}
\end{figure}

The authors in \cite{Chen2018SequentialMicrogrids, Ding2022ANetworks} presented a detailed restoration model considering the distribution system's bus-block structures and connecting switches (as shown in Fig. \ref{fig_example_DN}). Furthermore, the authors in \cite{Liu2021CollaborativeDERS} enhanced the restoration model of DS accounting the flexibility of behind-the-meters DERs. Another enhancement of restoration model is presented in \cite{Arif2022SwitchingRestoration} by considering capabilities and operational limits of different switching devices of DS. In \cite{Chen2019TowardRestoration}, the authors propose variable time step mixed-integer linear programming models to incorporate switches operating in multiple timescale while blackstarting DS. However, the models in \cite{Chen2018SequentialMicrogrids, Ding2022ANetworks,Liu2021CollaborativeDERS,Arif2022SwitchingRestoration,Chen2019TowardRestoration} are designed for blackstart generators with a very large energy capacity (e.g., diesel or thermal generators) and loads without cold load pick-up (CLPU) effect. Considering blackstart units with finite energy capacity such as BES-based GFMIs will potentially lead to different solution and challenges, which were not sudied in these above literature. The study in \cite{Gao2022DecentralizedResources} incorporates GFMI/GFLI models in DS restoration and provides the percentage of load pick up schedules in each buses; however, this approach is not practical as most DS loads are not dispatchable. Moreover, all these works neglect frequency security constraints, which are crucial for ensuring the stability of microgrids during blackstart operations.
\par
A sequential restoration of loads in the DS, forming multiple microgrids without exceeding maximum frequency nadir limits in each restoration sequence, is studied in \cite{Che2019AdaptiveConditions}. This is one of the pioneering works where the frequency nadir limit is estimated without the need to simulate the dynamic model of the DS with blackstart generators. However, a slight oversight of the work is that the microgrid also needs to satisfy the RoCoF and quasi-steady-state (QSS) frequency limits not only the frequency nadir for secure operation \cite{Nerc_reliability}. A restoration model incorporating blackstart GFMIs and non-blackstart GFLIs with frequency dynamic constraints is developed in \cite{Zhang2021AConstraints}. This model requires assistance from dynamic simulations to verify whether the frequency nadir exceeds secure limits when picking up loads. Although accurate, relying on simulations can be computationally expensive for large-scale restoration planning problems. The authors in \cite{Liu2023UtilizingRestoration} establish GFMI and GFLI models with a p-f droop-based relationship to estimate QSS frequency when restoring loads. However, frequency security constraints related to RoCoF and frequency nadir are ignored. All the above works neglect the synchronization of islands with the transmission grid (TG), resulting in multiple islanded microgrids forming at the end of the restoration process. However, the continuous operation of these microgrids beyond the restoration period cannot be guaranteed if the blackstart units are battery energy storage (BES)-based and the non-blackstart units are renewable-based. To address this issue, synchronization must be an integral part of the restoration process, a consideration currently missing in the literature. Additionally, synchronization among islands facilitates resource sharing and enables a faster connection with the TG. The significance of synchronization for a resilient distribution system (DS) is advocated in \cite{Du2022Black-StartMicrogrids}, where the synchronization is conducted at the end of the restoration process. However, their work has not optimized the energizing and synchronizing switching sequences.
\par
Hence, this paper proposes a holistic bottom-up blackstart and load restoration planning framework leveraging BES-based GFMIs and renewable GFLIs. The proposed framework initiates blackstart with multiple GFMIs, sequentially expands the boundaries of islanded microgrids while establishing cranking paths to GFLIs, synchronizes microgrids to form larger islands, and finally synchronizes with the TG to complete the restoration process. To incorporate synchronizing decisions in the problem, we first integrate virtual synchronous generator (VSG) control features into the GFMI. By leveraging VSG parameters (such as inertia constant and p-f droop), we establish and validate the GFMI's frequency response to cold load pick-up (CLPU) in terms of quasi-steady-state (QSS) frequency, Rate-of-Change-of-Frequency (RoCoF), and frequency nadir. This approach requires only the system's initial and final QSS points for frequency response estimation, making it suitable for optimization problems and eliminating the need for dynamic simulation assistance. Additionally, we model GFMI frequency adjustment to facilitate synchronization. Secondly, we model synchronizing switches, establishing conditions for their operation that align with the branch flow model of the DS. A dynamic radiality constraint is designed to support synchronization by allowing GFMIs to change the status of their root nodes to non-root nodes. Finally, all these models, along with GFLI dynamics, CLPU considerations, energizing switches, and the switching structure of the DS, are integrated with the linearized power flow model to develop a mixed-integer quadratic constrained problem for optimizing blackstart and load restoration. We validated the proposed framework using an unbalanced IEEE-123-bus system, considering cases with two and four GFMIs under various scenarios of TG recovery instants. 
To this end, the technical contributions of this work can be summarized as:\begin{itemize}
    \item Model a DERs-aided bottom-up blackstart framework that incorporates the frequency response of VSG-based GFMIs. Validate the GFMI's estimated frequency response to load pick-up events by using a dynamic model of the DS and GFMIs. 
    \item Model synchronizing switches (SSWs) with conditions that align with the branch flow-based restoration model. Optimize both the energizing and synchronizing switches efficiently in the blackstart and restoration process.
    \item Develop dynamic radiality constraints to support synchronization and allow microgrids to expand and integrate their boundaries during blackstart process.
\end{itemize}  


\section{Holistic formulation of blackstart and load restoration process}
In this section, we formulate an optimization problem for bottom-up blackstart and load restoration of the DS leveraging DERs. Our approach involves utilizing multiple BES-based GFMIs for blackstart and extending the cranking paths according to the DS switching structure to energize GFLIs while maximizing the cold load pickup without violating capacity and frequency security limits. Furthermore, we synchronize the microgrids to support each other and later with the TG upon its recovery, ensuring continuous operation beyond the studied restoration horizon. Our objective is to model the complete restoration process and determine the sequence of energizing and synchronizing switching decisions for a given TG recovery instant.
\subsection{Modeling BES-based grid-forming inverter (GFMI)}
\begin{figure}[t]
    \centering
    \includegraphics[width=\linewidth]{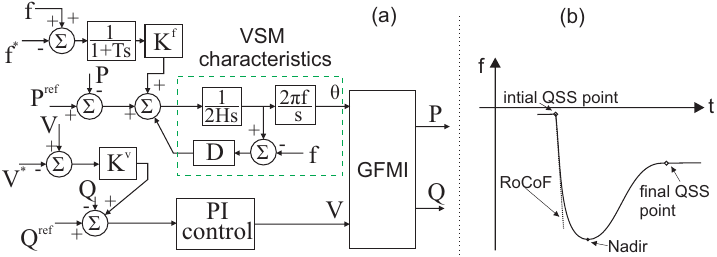}
    \vspace{-5mm}
    \caption{(a) Control diagram and (b) dynamic frequency response of a VSG.}
    \label{fig_vsm_ctrl}
\end{figure}
We consider all GFMIs to be 3-$\phi$ Virtual Synchronous Generator (VSG) designed with inertia constant (H) and damping constant (D) to emulate the synchronous generator's characteristics. A VSG operates with a voltage-frequency ($V-f$) control mode with a suitable frequency and voltage droops for active and reactive power sharing between other VSGs \cite{Ambia2021}.
\par
At normal operating conditions, GFMI regulates its terminal voltage and frequency close to nominal values, allowing them to change with respect to voltage and frequency droop gains ($K^v$ and $K^f$), as shown in Fig. \ref{fig_vsm_ctrl}a. To comply with the branch-flow model of the distribution system, we will refer to the square of voltage magnitude ($v = |V|^2$) as the control variable for GFMIs. Hence, the voltage and frequency of the GFMI at QSS are defined for all GFMI buses in $\mathcal{R}$ as:
\begin{align}
     v_{i,n,t} &= (V^*)^2 + \Delta v^{cc}_{i,t} \quad \forall i\in \mathcal{R}, n\in \Phi, \label{eqn_gfmi_vol}\\
    f_{i,t} &= f^*\Big(1-\frac{\sum_{n\in\Phi}{p^{ES}_{i,n,t}}}{S_i^{rat}(D_i+K^f_i)}\Big)  \quad \forall i\in \mathcal{R}.\label{eqn_gfmi_fre}
\end{align}
Here, $\Phi$ is a set of phases in bus i. $V^*$ and $\Delta v^{cc}$ are a nominal voltage reference and incremental voltage appearing due to the voltage droop. The derivation of (\ref{eqn_gfmi_vol}) is discussed in Appendix \ref{app_gfmi_vol}. Furthermore, (\ref{eqn_gfmi_fre}) is defined in \cite{Liu2024}, where $f^*$ is the nominal frequency set-point, $p^{ES}$ is the output active power, and $S^{rat}$ is the rated capacity of the inverter. GFMI has operational constraints over voltage, frequency, and power output, and are defined as:
\begin{align}
    &(0.95V^*)^2 \le v_{i,n,t} \le (1.05V^*)^2, \quad \forall i \in \mathcal{R}, n \in \Phi\\
    &\lfloor f^{qss}\rfloor \le f_{i,t}\le \lceil f^{qss} \rceil, \quad \forall i \in \mathcal{R} \\
    &\max_{n\in\Phi}\{(p^{ES}_{i,n,t})^2 +(q^{ES}_{i,n,t})^2\} \le \big(\tfrac{1}{3}S_i^{rat}\big)^2 \quad  \forall i \in \mathcal{R} \label{eq_gfmi_loadability}
\end{align}
Here, $\lfloor f^{qss}\rfloor$ and $\lceil f^{qss} \rceil$ are upper and lower boundaries of the frequency at quasi-steady-state. The thermal limit of GFMI (\ref{eq_gfmi_loadability}) has been adopted from \cite{Kim2022VoltageInverters}, where $q^{ES}$ refers to reactive power at each phase.Constraint (\ref{eq_gfmi_loadability}) signifies that once any phase of a GFMI reaches its thermal limit (i.e., $\tfrac{1}{3}S_i^{rat}$), the remaining two phases cannot be further loaded. BES-based GFMI will have an additional constraint on the state of charge ($SoC$), which is expressed as: 
\begin{align}
    SoC_{i,t} &= SoC_{i,t\text{-}1} + \frac{1}{C_i}\;\sum_{n\in\Phi}\{\Delta p^{ES}_{i,n,t}\} \Delta t, \quad \forall i \in \mathcal{R} \label{eq_soc}\\
    \text{where, } &\Delta p^{ES}_{i,n,t}=p^{ES}_{i,n,t} - p^{ES}_{i,n,t\text{-}1} \nonumber
\end{align}
Note that charging and discharging efficiencies are neglected in (\ref{eq_soc}) as the main focus during blackstart is to restore the loads more efficiently than considering the battery losses. $C_i$ and $\Delta t$ are battery capacity and a time-step, respectively.

\subsubsection{Constraining dynamic frequency responses to comply with security limits:}
When a GMFI picks up loads, the frequency initially declines abruptly during the first few seconds. After this initial drop, the frequency is regulated by droop control to bring it back within the QSS limits. This dynamic frequency response between the initial and final QSS points is shown in Fig. \ref{fig_vsm_ctrl}b. GFMI is required to comply its dynamic frequency response in terms of RoCoF and frequency-nadir ($f^{nad}$) for secure operation defined by NERC \cite{Nerc_reliability}. We leverage the work in \cite{Liu2024} to estimate RoCoF and$f^{nad}$ of VSG based on initial and final QSS operating points as:
\begin{align}
    &\dot{f}_{i,t} = \frac{\sum_{n\in\Phi}{\Delta P^{ES}_{i,n,t}}}{2S_i^{rat}H_i} \label{eqn_rocof}\\
    &f^{nad}_{i,t} = \Big[f_{i,t}-f^*\frac{\sum_{n\in\Phi}{\Delta P^{ES}_{i,n,t}}}{S_i^{rat}(D_i+K_i^f)} (1+\gamma_i)\Big] \label{eqn_fnadir}
\end{align}
Detailed calculation of $\gamma$ for VSGs is discussed in Appendix \ref{app_gamma}. Note that RoCoF and $f^{nad}$ for each VSG depend on its active power output either when operating autonomously or in parallel. Finally, we impose a security constraint on the dynamic frequency responses as:
\begin{align}
    &RoCoF^{min}\le\dot{f}_{i,t} \le RoCoF^{max} \quad \forall i \in \mathcal{R},\\   
    &f^{min}\le f_{i,t}^{nad}\le f^{max}\quad \forall i \in \mathcal{R}.
\end{align}

\subsection{Modeling synchronization of GFMIs}
GFMI needs to adapt its frequency set-points to allow synchronization with other GFMIs or the TG \cite{Rathnayake2022}. This is achieved by allowing perturbation in frequency ($\Delta f^{*}$) of GFMI while performing synchronization. Hence, (\ref{eqn_gfmi_fre}) is defined with synchronizing decisions as:
\begin{align}
    f_{i,t} &= f^*\Big(1-\frac{\sum_{n\in\Phi}{p^{ES}_{i,n,t}}}{S_i^{rat}(D_i+K_i^f)}\Big) +  \delta_{i,t} \Delta f_{i, t}^{*}. \label{eqn_fqss2}
\end{align}
Here, $\delta$ is a binary variable that is set to 1 only during the synchronizing period. Equation (\ref{eqn_fqss2}) allows frequency perturbation in  a GFMI attempting to synchronize, while keeping the frequencies of other GFMIs intact. This is like master-slave synchronization approach which can be more restrictive \cite{Lee_converters}. Hence, we implement co-operative synchronization approach, where all GFMIs decide the suitable synchronizing frequency, as:
\begin{align}
    f_{i,t} &= f^*\bigg(1-\frac{\sum_{n\in\Phi}{p^{ES}_{i,n,t}}}{S_i^{rat}(D_i+K_i^f)}\bigg) + \bigg(\sum_{b\in\mathcal{R}}\delta_{b,t}\bigg) \Delta f_{i, t}^{*}. \label{eqn_fqss3}
\end{align}
The $\delta$ is interlinked with the network radiality constraint defined in section \ref{sec_rad_sync}. Note that the bilinear term in (\ref{eqn_fqss2}) and (\ref{eqn_fqss3}) can be reformulated into convex constraints using McCormick envelopes \cite{CASTRO2015300}.

\begin{figure}[t]
    \centering
    \includegraphics[width=0.75\linewidth]{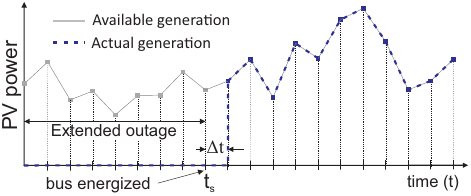}
    \vspace{-3mm}
    \caption{Energization of grid-following DERs such as PV.}
    \label{fig_grid-following}
\end{figure}
\subsection{Modeling grid-following PV inverters}
All 3-$\phi$ and 1-$\phi$ PV inverters are operated in grid-following mode, meaning they can come online only after the bus they are connected to is energized. If $y^B_{b,t}, b\in\mathcal{B}_{PV}$  are binary variables representing the energization status of buses where PVs are connected, then the grid-following nature of PV inverters can be modeled as:
\begin{align}
    P_{b,t}^{PV} = \bar{P}_{b,t}^{PV} \: y^B_{b,t\text{-}1}\hspace{2cm}\forall b\in \mathcal{B}_{PV} \label{eqn_PV_gen}
\end{align}
Here, $ P_{b,t}^{PV}$ and $\bar{P}_{b,t}^{PV}$ are actual and predicted PV power generation. Note that we multiply PV generation in (\ref{eqn_PV_gen}) by $y^B_{b,t\text{-}1}$ rather than $y_{b,t}^{B}$ due to consideration of an intentional time delay of one-time step. The typical inverter intentional time delay of grid-following renewable generation is 2-15 minutes, as recommended by IEEE 1547-2018. The impact of intentional time delay in the output PV power is illustrated in Fig. \ref{fig_grid-following}. The reactive power injections are assumed to be controllable, and their capabilities are defined to follow IEEE1547-2018 standards as:
\begin{align}
   -0.484 \lceil{P}\rceil_b^{PV} y^{B}_{b,t\text{-}1}\le Q_{b,t}^{PV}\le 0.484 \lceil{P}\rceil_b^{PV} y^{B}_{b,t\text{-}1}.
\end{align}
Here, $\lceil{P}\rceil_b^{PV}$ is the rated generation capacity of PV located at bus $b$.

\begin{figure}[t]
    \centering
    \includegraphics[width=0.75\linewidth]{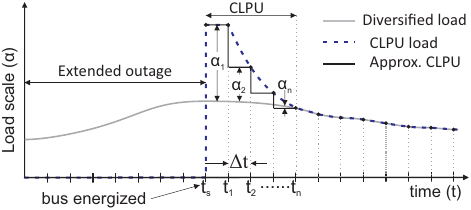}
    \vspace{-3mm}
    \caption{Cold load pickup model after an extended outage.}
    \label{fig_clpu}
\end{figure}

\subsection{Modeling cold load pick up (CLPU)}

Black starting a distribution system requires picking up dead loads, leading to higher power demand compared to its diversified form. The initial elevated demand gradually decreases with time in a non-linear way and finally follows the diversified demand, as illustrated in Fig. \ref{fig_clpu}. This phenomenon is called the CLPU effect \cite{gazijahani2022parallel}. 
\par
In this work, to eliminate the non-linearity introduced by the CLPU effect, a staircase modeling approach is developed, as shown in Fig. \ref{fig_clpu}. The developed approach captures the CLPU effect by introducing CLPU coefficients $\{\alpha_{1},\alpha_{2},\ldots,\alpha_{n}\}$, which progressively diminishes over time. Actual demand is then modeled as the product of their diversified load and the respective coefficient. 
\par
Generally, a distribution system has many hard-wired loads and few switchable loads. Hard-wired loads are non-switchable and are picked up upon energization of the bus to which they are connected. In contrast, switchable loads can be picked up in a controlled way. We assume switchable and non-switchable loads do not co-exist in the same bus for brevity. We represent buses containing switchable load by $\mathcal{B}_{CL}$ and non-switchable load by $\mathcal{B}_{NL}$, and assume  $\mathcal{B}_{CL}\cap \mathcal{B}_{NL} = \{\}$. Depending upon the status of the load $y_{b,t}^D$, the power demand of the switchable or non-switchable loads can be defined as:
\begin{align}
    P_{b,t}^{D} =& \bar{P}^{DL}_{b,t} \: \Big(\alpha_1 (y^{D}_{b,t}-y^{D}_{b,t\text{-}1})+\alpha_2(y^{D}_{b,t\text{-}1}-
    y^{D}_{b,t\text{-}2})+\nonumber\\&\alpha_3 (y^{D}_{b,t\text{-}2} - y^{D}_{b,t\text{-}3}) + y^{D}_{b,t}\Big) \quad \forall b \in \mathcal{B}_{CL}\cup \mathcal{B}_{NL},\label{eq_clpu_critical}\\
    Q^{D}_{b,t} =&  P_{b,t}^{D}\;tan(\theta_b^{D})\quad \forall b \in \mathcal{B}_{CL}\cup \mathcal{B}_{NL}.
\end{align}
Here, $\bar{P}_{b,t}^{DL}$ is forecasted diversified load. The terms within the bracket in (\ref{eq_clpu_critical}) define the approximated CPU effect. If the load at bus $b$ is non-switchable, then both the bus and load will have the same status, which is defined as:
\begin{align}
    y^{D}_{b,t}=y^{B}_{b,t}\quad\forall b \in \mathcal{B}_{NL}
\end{align}

\subsection{Modeling switching structure of distribution system}
A distribution system comprises a cluster of buses, referred to as bus-block \cite{Zhang2021AConstraints}, that are interconnected by switches, as shown in Fig. \ref{fig_example_DN}. During a blackstart, these bus blocks are energized sequentially with the help of blackstart resources by closing the switches one after another. It is essential to model the switching constraints to generate technically feasible switching sequences, which are discussed below.
\begin{figure}[t]
    \centering
    \includegraphics[width=\linewidth]{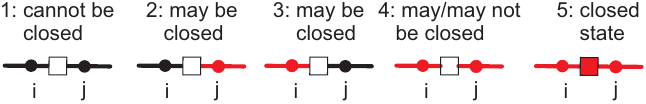}
    \vspace{-0.65cm}
    \caption{Possible switch states while blackstarting.}
    \label{fig_sw_state}
\end{figure}
\subsubsection{Modeling switches}
To integrate with the power flow model in section \ref{sec_pf}, we will represent all the switches by a negligible-impedance line $(i,j)$ where $i,j \in \mathcal{B}$. A switch may be in five states shown in Fig. \ref{fig_sw_state} while black-starting the system. In state 1, switching is infeasible because a switch requires cranking power to operate remotely. Switching is feasible at states 2 and 3 while switching at state 4 depends on the switch's capability. Only a switch with synchronizing capability can turn on at state 4. Hence, we will consider two types of switches: (a) energizing switches (ESW) and (b) synchronizing switches (SSW), and model their switching constraints. The switching constraints at all five states are shown in Table \ref{tab_sw_condition} for both the ESW and SSW, where $\Delta y_{ij,t}^L = y_{ij,t}^L - y_{ij,t\text{-}1}^L$.

\begin{table}[t]
  \centering
  \caption{Constraints for energizing and Synchronizing switches $(i,j)$}
      \begin{small}
  \vspace{-0.2cm}
  \renewcommand{\arraystretch}{1.3}
    \begin{tabular}{c|c|c|c|c|c}
    \toprule
      \multirow{2}{*}{state} &  \multirow{2}{*}{$y^B_{i,t\text{-}1}$}     &   \multirow{2}{*}{$y^B_{j,,t\text{-}1}$} &  \multirow{2}{*}{$y^L_{ij,t\text{-}1}$} & \multicolumn{2}{c}{$\Delta y^L_{ij,t}$  constraints}\\
       \cline{5-6}      &    &  &      & {ESW}     & SSW \\
    \hline
    1 & 0     & 0     & 0 & \multicolumn{2}{c}{$\Delta y^L_{ij,t}=0$} \\
    \hline
    2 & 0     & 1     & 0 & \multicolumn{2}{c}{$\Delta y^L_{ij,t}\le 1$}\\
    \hline
    3 & 1     & 0     & 0 & \multicolumn{2}{c}{$\Delta y^L_{ij,t}\le 1$}\\
    \hline
    4 & 1     & 1 & 0   &  $\Delta y^L_{ij,t}= 0$& $\Delta y^L_{ij,t}\le 1$\\
    \hline
    5 & 1 & 1& 1 & \multicolumn{2}{c}{$\Delta y^L_{ij,t}= 0$}\\
    \bottomrule
    \end{tabular}%
  \label{tab_sw_condition}%
  \end{small}
\end{table}%
 
\paragraph{Energizing switch (ESW)}
Energizing switches (e.g., breakers, reclosers, or tie-switches) are turned on to power the inactive portion of the DS. However, they cannot close the switch to connect two active buses, as these may have different phases and frequencies. We define the set of all ESW by $\mathcal{L}^{ESW}$. All the constraints for every state of ESW in Table \ref{tab_sw_condition} can be succinctly defined for all $(i,j)\in \mathcal{L}^{ESW}$ as:
\begin{align}
    &y^L_{ij,t} \le y^B_{i,t\text{-}1} + y^B_{j,t\text{-}1} \\
    &\Delta y^L_{ij,t} \le 2 - y^B_{i,t\text{-}1} - y^B_{j,t\text{-}1}, \quad \Delta y^L_{ij,t} \ge 0
\end{align}

\paragraph{Synchronizing switch (SSW)}
Synchronizing switches are special switches equipped with Intelligent Electronic Devices (IEDs) with the additional functionality of monitoring voltages across the switch at a fast rate to facilitate synchronization of the grids. They are also referred to as smart switches in \cite{Du2022Black-StartMicrogrids}. We represent all SSWs by a set $\mathcal{L}^{SSW}$. Similar to ESW, switching constraints for all $(i,j) \in \mathcal{L}^{SSW}$ can be defined as:
\begin{align}
    y^L_{ij,t} \le y^B_{i,t\text{-}1} + y^B_{j,t\text{-}1}, \quad \Delta y^L_{ij,t} \ge 0
\end{align}
To close an SSW, it is significant to match the voltage and angle of buses $i$ and $j$ to ensure secure connection of grids. Imposing this constraint is challenging when using the branch-flow model (outlined in Section \ref{sec_pf}), as it does not use node voltages in polar form. However, we can satisfy this constraint by ensuring negligible (ideally zero) power flow across the SSW when it is switched on and then relaxing the power flow afterward. This will require identifying the instant of closing SSW. We know when SSW is switched ON, $\Delta y_{ij,t}^L \rightarrow 1$. However,  this is not sufficient as this condition is true even for energizing switching. Additionally, when SSW is switched ON, the condition $(y_{i,t\text{-}1}^B+y_{j,t\text{-}1}^B-y_{ij,t}^L)\rightarrow 1$, which does not hold for energizing switching. Hence, the synchronizing instant can be defined by the multiplication of two conditions, as:
\begin{align}
    z_{ij,t}^L = \Delta y_{ij,t}^L (y_{i,t\text{-}1}^B+y_{j,t\text{-}1}^B-y_{ij,t}^L)
\end{align}
Utilizing $z_{ij,t}^L$, the synchronizing constraint can be defined for all $(i,j)\in\mathcal{L}^{SSW}$ as:
\begin{align}
    -(1-z_{ij,t}^L)M-\epsilon \le P_{ij,t}\le \epsilon + (1-z_{ij,t}^L)M\\
    -(1-z_{ij,t}^L)M-\epsilon \le Q_{ij,t}\le \epsilon + (1-z_{ij,t}^L)M
\end{align}
Here, $M$ and $\epsilon$ are big and small numbers. 
\par 
When an ESW or a SSW $(i,j)$ is closed, the frequencies at buses $i$ and $j$ should match; otherwise they may operate at different frequencies. This constraint can be defined for all $(i,j)\in \mathcal{L}^{ESW}\cup\mathcal{L}^{SSW} $ as:
\begin{align}
    -(1-y_{ij,t}^L)M-\epsilon \le f_{i,t} - f_{j,t}\le \epsilon + (1-y_{ij,t}^L)M
\end{align}

\subsubsection{Bus blocks energizing constraints}
\begin{figure}[t]
    \centering
    \includegraphics[width=0.85\linewidth]{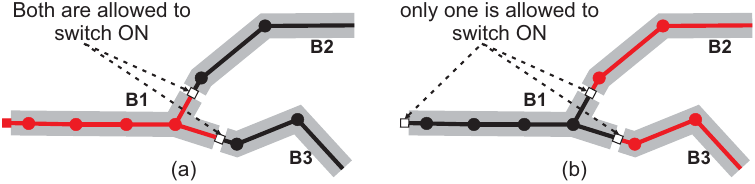}
    \vspace{-0.45cm}
    \caption{(a) Multiple switches associated with active bus block (e.g, B1) can be switched ON. (b) Only one switch associated with inactive bus block (e.g., B1) can be switched ON.}
    \label{fig_sw_BB}
\end{figure}

In a multi-configurable distribution system, a bus block is segregated by multiple switches (for reference, see Fig. \ref{fig_example_DN}) and, hence, can be energized by closing any of them. Additionally, once a bus block is energized, it is not de-energized. We represent a set of bus blocks of DS by $\mathcal{M}$ and the switches associated with each bus block $m ( m\in \mathcal{M})$ by a set $\mathcal{W}(m)$. Hence, the energizing constraints of a bus block are modeled as follows.
\begin{align}
    y^{BB}_{m,t} &\ge y^{L}_{ij,t} \quad \forall \{m\in\mathcal{M}|(i,j)\in \mathcal{W}(m)\}\\
    y^{BB}_{m,t} &\ge y^{BB}_{m,t\text{-}1}
\end{align}
Here, $y^{BB}_{m,t}$ is a binary variable representing its activation status. Furthermore, an energizing of bus block will activate its lines and buses, which is modeled as:
\begin{align}
    y^L_{ij,t} &= y^{BB}_{m,t} \quad \forall\{m\in\mathcal{M}| (i,j) \in \mathcal{L}(m)\}\\
    y^B_{i,t} &= y^{BB}_{m,t} \quad \forall \{m\in\mathcal{M}|i\in \mathcal{B}(m)\} \label{eq_enes1}
\end{align}
Here, $\mathcal{L}(m)$ and $\mathcal{B}(m)$ represent the set of lines and buses associated with bus block $m$. To ensure the feasibility of switching, we allow multiple switches associated with an active bus block to be turned on, but only one switch for an inactive bus block. An inactive bus block is limited to only one switch to prevent energization from multiple active bus blocks, which may belong to different microgrids where synchronization will be necessary before merging them. This constraint is illustrated in Fig. \ref{fig_sw_BB} and can be define as:
\begin{align}
    \sum_{(i,j)\in \mathcal{W}(m)} \hspace{-1em}  y_{ij, t}^L - \hspace{-1em}\sum_{(i,j)\in \mathcal{W} (m)}  \hspace{-1em}y_{ij, t\text{-}1}^L \le M y_{m, t}^{BB} + 1 \quad \forall m \in \mathcal{M}.
\end{align}
Furthermore, the frequency measured at any bus within the bus block should have same value. This can be achieved with the help of following constraint defined for each segment $m$.
\begin{align}
    f_{b,t}= f_{m,t}\quad \forall b \in \mathcal{M}(m)
\end{align}

\subsection{Linear power flow model adaptive to network topology}\label{sec_pf}
The study conducted in \cite{Cheng2022} investigated a linearized power flow model for an unbalanced distribution system with all active lines and buses. However, the number of active lines and buses are changing while blackstarting. To adapt with active/inactive elements, we relax the power flow equations in \cite{Cheng2022} with the help of binary variables representing the active/inactive status of the lines ($y_{ij}^L$) and buses ($y_i^B$). We establish a power flow model by defining voltage equations for all lines ($i.e., \forall (i,j)\in \mathcal{L}$) as:

\begin{subequations}
\begin{align}
    &\boldsymbol{v}_{j,t} \le \boldsymbol{v}_{i,t}^{\Phi_{ij}}-
    2(\obar{\boldsymbol{R}}_{ij}\boldsymbol{P}_{ij,t}+ \obar{\boldsymbol{X}}_{ij}\boldsymbol{Q}_{ij,t}) + \boldsymbol{M}^{\Phi_{ij}} (1-y^L_{ij,t}) \label{eqn_dist_v}\\
    &\boldsymbol{v}_{j,t} \ge \boldsymbol{v}_{i,t}^{\Phi_{ij}}-
    2(\obar{\boldsymbol{R}}_{ij}\boldsymbol{P}_{ij,t}+ \obar{\boldsymbol{X}}_{ij}\boldsymbol{Q}_{ij,t}) - \boldsymbol{M}^{\Phi_{ij}} (1-y^L_{ij,t}) \label{eqn_dist_v1}\\
    &-\boldsymbol{M}^{\Phi_{ij}} y^L_{ij,t} \le \boldsymbol{P}_{ij,t}\le  \boldsymbol{M}^{\Phi_{ij}} y^L_{ij,t}\\
    &-\boldsymbol{M}^{\Phi_{ij}} y^L_{ij,t} \le \boldsymbol{Q}_{ij,t}\le  \boldsymbol{M}^{\Phi_{ij}} y^L_{ij,t}\\  
    &\boldsymbol{0.9025}^{\Phi_{ij}}\: y_{j,t}^B \le \boldsymbol{v}_{j,t} \le  \boldsymbol{1.1025}^{\Phi_{ij}}\: y_{j,t}^B
\end{align}\label{eq_pf_unbalanced}
\end{subequations}

Here,  $\boldsymbol{v}_{j,t}$ is a vector representing the square of voltage magnitude at each phase in the bus $j$; $\boldsymbol{v}_{i,t}^{\Phi_{ij}}$ is sub-matrix of  $\boldsymbol{v}_{j,t}$ containing the entries associated with $\Phi_{ij}$; $P_{ij}$ and $Q_{ij}$ are real and reactive power flow on line $(i,j)$; $\boldsymbol{M}$ is a user-defined large constant vector that helps to relax the power flow constraints (\ref{eq_pf_unbalanced}a)-(\ref{eq_pf_unbalanced}b) even when the line $(i,j)$ is unenergized (i.e. when $y_{ij,t}^L\rightarrow 0$). A multi-configurable distribution system is an undirected graph, particularly while black-starting, without prior information of parent-child relationships among its buses. Nonetheless, the relaxation in (\ref{eq_pf_unbalanced}c)- (\ref{eq_pf_unbalanced}d) allows bidirectional power flow, allowing us to initially treat it as a directed graph with assumed parent-child relationships. Finally,  (\ref{eq_pf_unbalanced}e) is a box constraint for $\boldsymbol{v}_{i,t}$ to regulate voltage magnitude between 0.95 to 1.05 p.u. if the status of the bus is active (i.e., when $y_j^B\rightarrow 1$), otherwise squeezed to 0. $\obar{\boldsymbol{R}}_{ij}$ and $\obar{\boldsymbol{X}}_{ij}$ are matrix parameters associated with line resistance and reactance, whose computations are shown in \cite{Cheng2022}.
\par
Neglecting the line losses, the power flow on line $(i,j)$ is written for all $(i,j)\in \mathcal{L}$ as:

\begin{subequations}
   \begin{align}   
    \boldsymbol{P}_{ij,t} =& \sum_{k\in N(j)}\boldsymbol{P}_{jk,t} - \boldsymbol{p}_{j,t}\\
    \boldsymbol{Q}_{ij,t} =&\sum_{k\in N(j)}\boldsymbol{Q}_{jk,t} - \boldsymbol{q}_{j,t}
\end{align} \label{eqn_net_power_balacne}
\end{subequations}

Here, $N(j)$ is a set child bus of bus $j$; $\boldsymbol{p}_{j}/\boldsymbol{q}_{j}$ is a vector of net active/reactive power injection at bus $j$ such that generation is assumed positive and demand is negative.

\subsection{Modeling transmission grid outage and recovery}
A DS is disconnected from out-of-service TG during the blackstart and later reconnected when the TG is restored. We represent a set of transmission-distribution interconnection bus as $\mathcal{B}^{TG}$ and its active/inactive status by a binary parameter $y_{b,t}^{TG}$. Note that $y_{b,t}^{TG}$ is an input parameter, whose value is  set to 0 when the TG is out-of-service and 1 when it is active. Hence, the TG is modeled for all $b\in \mathcal{B}^{TG}$ as:
\begin{subequations}
   \begin{align}
    &(P^{TG}_{b, t})^2 + (Q^{TG}_{b,t})^2 \le (SS^{rat})^2,\\
    &1.0y_{b,t}^{TG}\ge v_{b,t} \ge 1.0y_{b,t}^{TG},\\
    & 60y_{b,t}^{TG}\ge f_{b,t} \ge 60y_{b,t}^{TG}
\end{align} \label{eqn_ss_constraints}
\end{subequations}

\subsection{Dynamic radiality constraint driven synchronization} \label{sec_rad_sync}
Radiality constraint must be maintained at every switching sequence while blackstarting the DS. In \cite{wang2020radiality}, radiality constraint is developed for static case, where the network size and number of root buses are fixed. However, in the black-starting problem, the radiality constraint has to be satisfied adapting to activation status of the buses and lines. A bus with an active grid-forming source (e.g., GFMI and TG) can be considered a root bus. The number of active root bus defines the number of microgrids that can be formed while maintaining the tree or radial structure. Hence, the dynamic radiality constraints are defined as:
\begin{subequations}
\begin{align}
    \sum_{(i,j) \in \mathcal{L}} y^L_{ij,t} &= \sum_{b\in\mathcal{B}} y_{b,t}^B + \sum_{b\in\mathcal{B}^{TG}} y_{b,t}^{TG}- R_t \label{eq_rad4}\\
    R_t &= \sum_{b \in \mathcal{R}} y^{ES}_{b,t} + \sum_{b\in\mathcal{B}_{TG}} y_{b,t}^{TG}\\
    y^{ES}_{b,t}&\le y^{ES}_{b,t\text{-}1}\quad \forall b \in \mathcal{R}
\end{align}\label{eqn_rad_const}
\end{subequations}

In (\ref{eqn_rad_const}a), the activation status of substation bus ($y_{b,t}^{TG}$) and other buses ($y_{b,t}^B$) have been considered separately because $y_{b,t}^{TG}$ is an input parameter to simulate various outage duration of TG, whereas $y_{b,t}^B$ are part of decision variables. Furthermore, $R_t$ is also the decision variable defining the number of root buses in (\ref{eqn_rad_const}b). We allow GFMI buses to be non-root buses but do not permit them to become root buses again, by defining (\ref{eqn_rad_const}c). Although this formulation decide the optimal number of microgrids along the black-starting and load restoration process, it is essential to coordinate frequency perturbation decisions ($\Delta f_{b,t}^{*})$ in GFMIs, to match frequencies. This is achieved by activating $\delta_{b,t}$ at synchronizing instant, by defining a constraint as:
\begin{align}
    y_{b,\tau}^{ES} = 1-\sum_{t\in[t_o,\tau]} \delta_{b,t} \quad \forall \tau \in \mathcal{T}, \forall b \in \mathcal{R} \label{eqn_coor_syn}
\end{align}
Constraint (\ref{eqn_coor_syn}) signifies that whenever a GFMI at bus $b$ change its status from root to non-root bus, $\delta_{b,t}$ set to 1, otherwise 0. Here, $t_o$ is start time of balckstart and $\mathcal{T}$ is the time horizon of study.

\subsection{Optimization model}
We propose a mixed-integer model for black start and load restoration planning where the primary decision variables are root bus status of GFMIs ($\boldsymbol{y}^{B}_{b,t}, b\in\mathcal{R},t\in\mathcal{T}$), ESW/SSWs' statuses ($\boldsymbol{y}^{L}_{ij,t}, (i,j)\in\mathcal{L}^{ESW}\cup\mathcal{L}^{SSW},t\in\mathcal{T}$), bus blocks' statuses ($\boldsymbol{y}^{BB}_{m,t}, m\in\mathcal{M}$), and status of switchable loads ($\boldsymbol{y}^{NL}_{b,t}, b\in\mathcal{B}^{NL},t\in\mathcal{T}$). Other decision variables pertaining to the activation of PVs, non-switchable loads, lines, and buses are associated with the bus block's status ($y^{BB}_{m,t}$). For convenience, we associate all the decision variables into vector $\boldsymbol{x}$ as $\boldsymbol{x} = [\boldsymbol{y}^{B}_{b,t}, \boldsymbol{y}^{L}_{ij,t}, \boldsymbol{y}^{BB}_{m,t}, \boldsymbol{y}^{NL}_{b,t}]^T$.

\subsubsection{Objective function}
The objective is to maximize the load restoration throughout the black start process. Hence, the objective is formulated as:

\begin{align}       
\max_{\boldsymbol{x}}{\sum_{t \in \mathcal{T}}\sum_{b \in  \mathcal{B}^{NL}\cup\mathcal{B}^{CL}} P_{b,t}^D \Delta t}.
\end{align}

\subsubsection{Associated constraints}
All the constraints pertaining to individual resources and networks defined in (\ref{eqn_gfmi_vol}) to (\ref{eqn_coor_syn}) are imposed on the optimization model. 
\subsubsection{Model Initialization}
All GFMI are self-started, considering them to be the root buses with full SoC. Statuses of all bus blocks and switches are considered inactive. We define these initial conditions as:
\begin{subequations}
    \begin{align}
y^{ES}_{b, t\rightarrow 0} &= 1, \;\; y^B_{i,t\rightarrow 0} = 1, \;\; SoC_{b, t\rightarrow 0}=1 \quad \forall i \in \mathcal{R}\\
y^{BB}_{m, t\rightarrow 0} &=0\quad \forall m \in \mathcal{BB} \\
y^{L}_{ij,t\rightarrow 0} &=0\quad \forall (i,j) \in \mathcal{L}^{ESW}\cup\mathcal{L}^{SSW}
\end{align}
\end{subequations}
\subsubsection{Modeling different synchronizing settings}\label{sec_syn_settings}
\begin{figure}[t]
    \centering
    \includegraphics[width=0.8\linewidth]{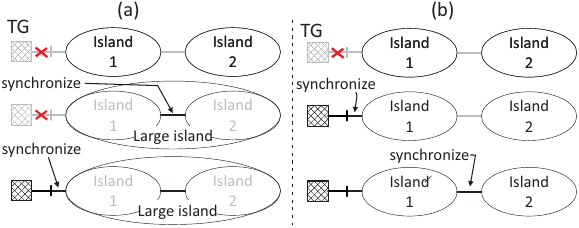}
    \vspace{-3mm}
    \caption{Synchronizing settings: (a) optimally synchronizing the islands while black starting to form a large island, which is later synchronized with TG; (b)  synchronizing islands one after another with TG after TG is available.}
    \label{fig_sync_settings}
\end{figure}

The black start and load restoration problem has traditionally been examined without considering the impact of synchronizing switching and generally ends up with the restoration forming multiple islands. Synchronization with the TG is essential to ensure the continuity of service beyond the study period ($\mathcal{T}$). The above problem can be modeled in two different synchronizing settings, as shown in Fig. \ref{fig_sync_settings}.
 
\paragraph{Optimal switching for energizing and synchronizing}
This is our proposed setting, where we consider a number of root bus ($R_t$) as an optimization variable having lower and upper bounds of 1 and $|\mathcal{R}+\mathcal{B}_{ss}|$, respectively. Furthermore, both the energizing and synchronizing switching are optimally decided while black starting.

\paragraph{Switching for synchronizing after TG is available}
In this setting, a number of root buses ($R_t$) are fixed to the number of GFMIs, and only energizing switching is optimized. After a TG is available, it is reduced progressively step by step with a pre-defined reconnecting rule. A general rule is to synchronize the island near the TG at first and then keep on synchronizing its neighboring island until all the islands are connected to the grid, as shown in Fig. \ref{fig_sync_settings}b.  
\begin{figure}[t]
    \centering
    \includegraphics[width=\linewidth]{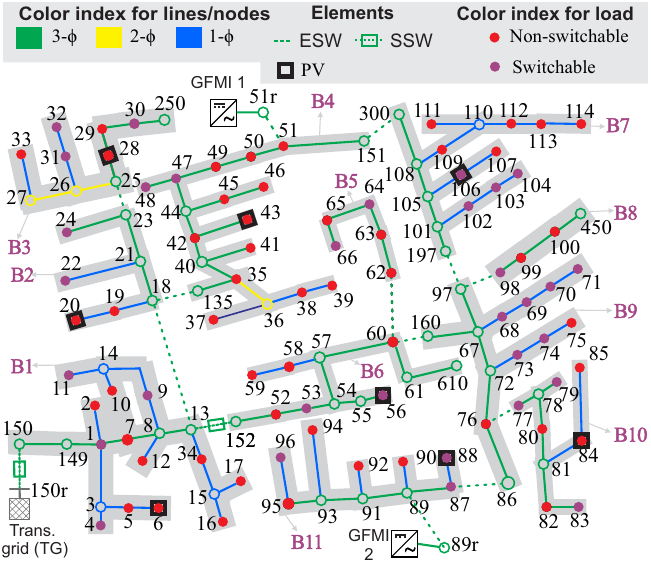}
    \vspace{-7mm}
    \caption{Modified IEEE 123 bus test distribution system.}\label{fig_IEEE_123bus}
\end{figure}
\begin{table}[t]
  \centering
  \caption{Location of SSWs and GFMIs in IEEE-123-bus test system }
  \begin{small}
    \begin{tabular}{|c|c|c|c|}
    \hline
    \multirow{2}{1.2cm}{Case studies} & \multirow{2}{2.8cm}{Location 
 :capacity (bus : MVA/ MWh)} & \multirow{2}{*}{{SSW locations}} \\
    & &\\
    \hline
      \multirow{2}{*}{2 GFMIs} & 51r : 2.45/4.6  &  \multirow{2}{*}{(150r, 150) , (13, 152)}\\
      & 89r : 2.65/3.42   &   \\\hline
     \multirow{4}{*}{4 GFMIs}   & 51r : 1.35/2.5   &   \multirow{4}{3cm}{(150r, 150) , (18, 135), (13, 18), (60, 160), (97, 197)}  \\
      & 89r : 1.5/2.35   &    \\
      & 29r : 0.95/1.17 &\\
      & 152r : 1.3/2.0 &\\
    \hline
    \end{tabular}%
  \label{tab_BES_ratings}%
  \end{small}
\end{table}%

\section{Results}
To analyze the performance of the proposed blackstart and load restoration framework, we employed the IEEE-123-bus test DS, which is an unbalanced network having a maximum load demand of 3.49 MW. As shown in Fig. \ref{fig_IEEE_123bus}, we segmented the entire network into 11 bus-blocks (B1 to B11) interconnected by either a ESW or SSW. All the switches are considered to be 3-$\phi$. Furthermore, 1-$\phi$/3-$\phi$ PVs are allocated throughout the network, with locations indicated by square-shaped boxes. A total of 965 kW of PVs are installed in this test system, which accounts for about 28\% of the maximum demand. Reactive power from PV inverters is assumed controllable while complying with IEEE 1547 standards.To represent a realistic situation, we assumed that 60\% of point loads are non-switchable (hard-wired) and the remaining 40\% are switchable. Their locations are shown by red and purple circles in Fig. \ref{fig_IEEE_123bus}. We will conduct case studies to demonstrate the advantages of the proposed framework, considering scenarios with two and four BES-based GFMIs. Locations and power/energy capacity of BES-based GFMIs for both the case studies are listed in Table \ref{tab_BES_ratings}.In both case studies, the aggregate power and energy capacity of the GFMIs are identical. Their total effective capacity (i.e., state of charge between 1 and 0.2) is sufficient to sustain all non-switchable loads for only three hours. Additionally, the locations of SSWs are shown in Table \ref{tab_BES_ratings}, while all other switches are ESWs.

\par 

Firstly, we will verify the accuracy of the estimated frequency responses of GFMIs using an RMS model of the DS in DigSILENT PowerFactory. Secondly, considering only two GFMIs, we will analyze, evaluate, and benchmark the performance of the proposed framework under various scenarios of TG restoration following an extended blackout. Lastly, we will summarize the performance of the proposed framework for four GFMIs.

\subsection{Verification of frequency responses of GFMIs}
\begin{figure}[t]
    \centering
    \includegraphics[width=\linewidth]{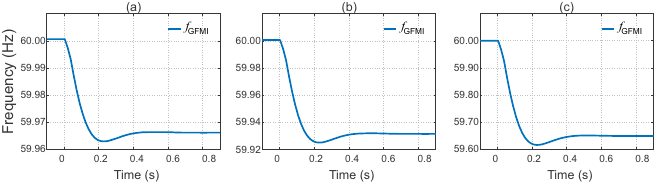}
    \vspace{-8mm}
    \caption{Transient frequency response of the GFMI under different pick-up loads. (a) 1 MW load pickup. (b) 2 MW load pickup. (c) 10 MW load pickup.}
    \label{fig_transientfrequency}
\end{figure}

We built a distribution network with virtual-synchronous-generator-based GFMI model in the DigSILENT PowerFactory and executed RMS simulation with load pickup events to evaluate the accuracy of  estimated frequency responses listed in (\ref{eqn_gfmi_fre}), (\ref{eqn_rocof}), and (\ref{eqn_fnadir}). The transient responses of the GFMI are recorded for 1, 2, and 10 MW load pickups, as show in in Fig. \ref{fig_transientfrequency}. This dynamic simulation provided actual values of $f^{qss}$, RoCoF, and $f^{nad}$, which are used as references to evaluate the accuracy of the estimations, as shown in Table \ref{table_frequencyvalidation}. The accuracy of the estimated frequency responses is above 92\%, which is commendable as they only utilize initial and final QSS points. This method is more favorable for large-scale planning problem as they eliminate the requirement of assistance from the complex dynamic simulation, proposed in \cite{Zhang2021AConstraints}.   

\begin{table}[t]
\centering
\caption{Validation of Estimated Frequency responses}\label{table_frequencyvalidation}
\begin{small}
\begin{tabular}{c c c c}\hline
    Pick-up load (MW) & 1 & 2 & 10 \\ \hline
    Measured $\Dot{f}$ (Hz/s) & -0.3529 & -0.7058 & -3.6106 \\
    Estimated $\Dot{f}$ (Hz/s) & -0.3780 & -0.7500 & -3.7500\\
    Accuracy of $\Dot{f}$ (\%) & 92.89 & 93.74 & 96.14\\
    Measured $f^{nad}$ (Hz) & 59.9629 & 59.9251 & 59.6161 \\ 
    Estimated $f^{nad}$ (Hz) & 59.9635 & 59.9272 & 59.6357\\
    Accuracy of $f^{nad}$ (\%) & 98.38 & 97.20 & 94.89 \\
    Measured $f^{qss}$ (Hz) & 59.9662 & 59.9316 & 59.6490 \\ 
    Estimated $f^{qss}$ (Hz) & 59.9666 & 59.9333 & 59.6666\\
    Accuracy of $f^{qss}$ (\%) & 98.82 & 97.51 & 94.99 \\\hline
\end{tabular}
\end{small}
\end{table}
\begin{figure}[t]
    \centering
    \includegraphics[width=\linewidth]{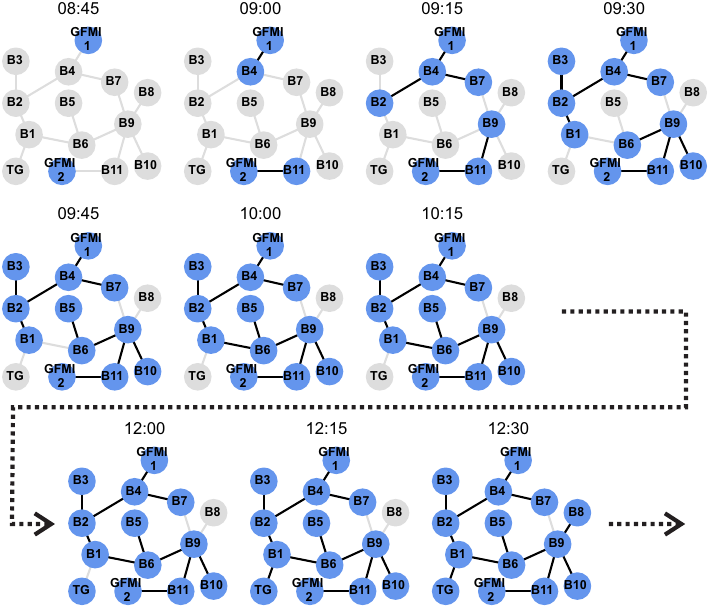}
    \vspace{-5mm}
    \caption{Evolution of cranking paths in the proposed method.}\label{fig_cs1_seq_seg_switch}
\end{figure}

\begin{figure}[t]
    \centering
    \includegraphics[width=\linewidth]{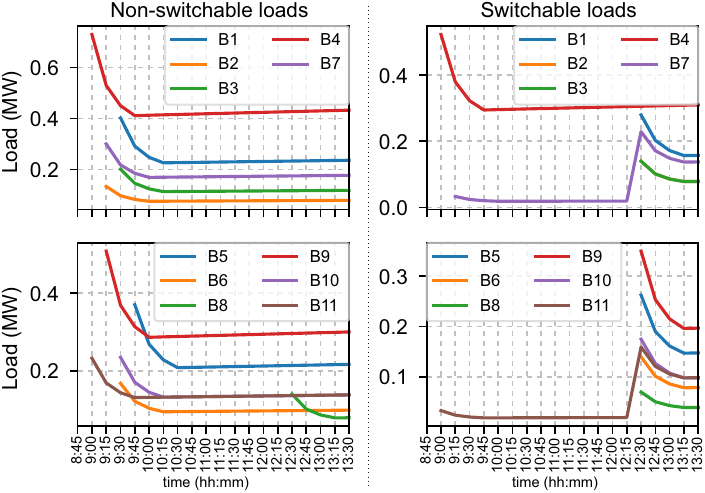}
    \vspace{-8mm}
    \caption{Restoration of loads in the proposed method.}\label{fig_cs2_load_restoraiton}
\end{figure}

\begin{figure}[t]
    \centering
    \includegraphics[width=\linewidth]{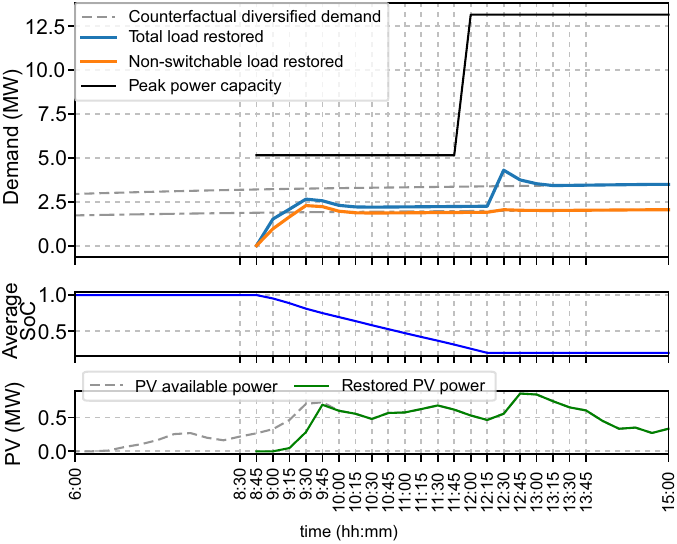}
    \vspace{-8mm}
    \caption{Performance of blackstart and load restoration in the proposed method.}\label{fig_cs2_performance}
\end{figure}
\subsection{Blackstart and load restoration with two GFMIs}
We will study the proposed framework under two synchronization settings, illustrated in Section \ref{sec_syn_settings}, for a scenario where a blackstart is initiated at 8:45 and the TG comes online at 12:00. Furthermore, we will compare restoration metrics for other restoration instants of TG. 
\subsubsection{Proposed method -- optimized energizing and synchronizing decisions:} 
In the proposed method, we optimize both energizing and synchronizing decisions to enhance the performance of blackstart and load restoration. The optimal sequence of establishing cranking paths obtained by the proposed method is shown in Fig. \ref{fig_cs1_seq_seg_switch}. Here, the blackstart initiates from two GFMIs at buses 51r and 89r at 8:45. Subsequently, the bus-blocks B4 and B11 are energized simultaneously by closing switches adjacent to GFMIs, leading to formation of two islanded microgrids. Furthermore at 9:15, bus-blocks B2 and B7 are energized in the first microgrid, while only B9 is energized in the second microgrid. The boundary of these two microgrids expands progressively at each time step while picking up loads and energizing PVs associated with the bus-blocks till 9:45. Note that at 9:45, only B5 is energized while B8 is not, due to the insufficient energy capacity of GFMI2. At 10:00, two microgrid are sychronized and connected by closing a SSW between B1 and B6. The distribution system remains in this configuration till the TG comes online at 12:00. At 12:15, the entire network is synchronized with TG, and B8 is picked up at 12:30. This is the final configuration at which DS is operated after the successful blackstart. Note that when the bus-blocks are energized, all non-switchable loads are picked up at the same instant, as seen in Fig. \ref{fig_cs2_load_restoraiton}. Depending upon the energy and power capacity of GFMIs, only few switchable loads are picked up while bus-blocks are energized. Many switchable loads are restored only after connecting with the TG, as shown in Fig. \ref{fig_cs2_load_restoraiton}. Note that the system experienced higher demand at the instant of load pickup, which decayed approximately in an exponential manner due to the CLPU effect.
\par
\begin{figure}[t]
    \centering
    \includegraphics[width=\linewidth]{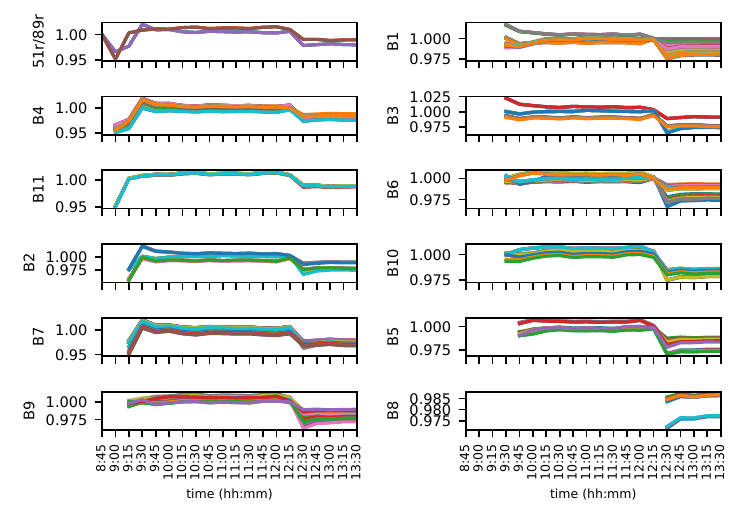}
    \vspace{-10mm}
    \caption{Nodes' voltage magnitude at bus-blocks in the proposed method.}\label{fig_cs1_voltage}
\end{figure}

Fig. \ref{fig_cs2_performance} illustrates the system restoration performance of the proposed method. The amount of restored loads increased with every switching decisions after the blackstart. Due to hard-wired connection, the non-switchable loads are restored quickly while forming the cranking paths to energize grid-following PVs. In contrast, the restoration of switchable loads can be deferred to prioritize expanding cranking paths and bringing more PVs online, thereby enhancing the overall restoration process. Hence, they are restored mostly after connecting with the TG, as shown in Fig. \ref{fig_cs2_performance}. The amount of aggregated PV generation is seen to increase in Fig. \ref{fig_cs2_performance} because of more PVs coming online while expanding the cranking paths. It can be observed from the SoC curve that the energy capacity of both the GFMIs are fully utilized to blackstart and restore the loads before TG comes online. The voltage at each node during the blackstart and load restoration process is maintained between 0.95 to 1.05 pu, as shown in Fig. \ref{fig_cs1_voltage}.

\begin{figure}[t]
    \centering
    \includegraphics[width=\linewidth]{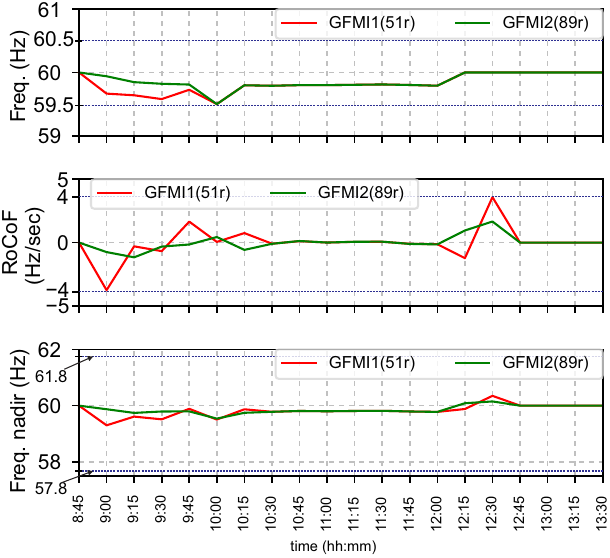}
    \vspace{-8mm}
    \caption{Frequency constraints of GFMIs in the proposed method}\label{fig_cs1_freq_const}
\end{figure}

All the frequency constraints for GFMIs are within an acceptable range, as shown in Fig. \ref{fig_cs1_freq_const}. When blackstart is initiated at 8:45, both GFMIs operate at 60 Hz. However, their frequencies drop independently due to drooping characteristics as loads are picked up, until they synchronize at 10:00. After synchronization, they share the same QSS frequencies. At 12:15, both GFMIs synchronize with the TG and operate at 60 Hz. RoCoF and $f^{nad}$ are metrics used to define the transient behavior of GFMI. Both metrics are within secure limits, as shown in Fig. \ref{fig_cs1_freq_const}.

\subsubsection{Benchmark method -- Optimal energizing decisions and rule-based synchronization.}
Existing research on blackstart and load restoration with DERs does not account for synchronization. Consequently, their approaches result in the formation of multiple islanded microgrids, raising concerns about how long these microgrids can be sustained given the limited energy capacity of blackstart units such as BES-based GFMIs. For a fair comparison with the proposed method, we will implement a rule-based synchronization (as discussed in Section \ref{sec_syn_settings}) of all the islanded microgrids once the TG is available after an extended blackout. Therefore, in this benchmark method, we are only optimizing the ESWs, while the SSWs are operated in a predefined sequence. 
\par
\begin{figure}[t]
    \centering
    \includegraphics[width=\linewidth]{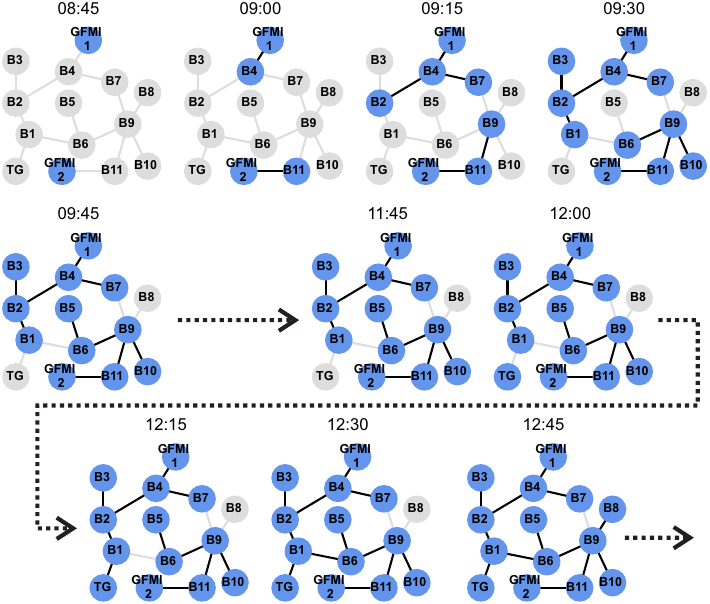}
    \vspace{-7mm}
    \caption{Evolution of cranking paths in the benchmark method.}\label{fig_cs2_seq_seg_switch}
\end{figure}
\begin{figure}[t]
    \centering
    \includegraphics[width=\linewidth]{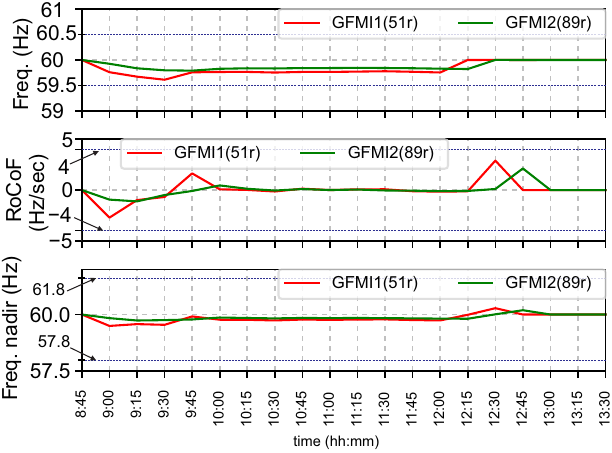}
    \vspace{-8mm}
    \caption{Frequency constraints of GFMIs in the benchmark method.}\label{fig_cs2_freq_const}
\end{figure}

The evolution of cranking paths with the benchmark method is depicted in Fig. \ref{fig_cs2_seq_seg_switch}. Unlike in the proposed method, the two islanded microgrids do not synchronize and remain islanded until the TG appears at 12:00. Subsequently, the microgrid near the TG is synchronized at 12:15, followed by the synchronization of the other microgrid at 12:30. Finally, the bus-block B8 is energized at 12:45. This configuration of the distribution system is maintained for further operations after a successful black start. The detailed results of benchmark method are intentionally not highlighted as they closely corroborate with the proposed method. However, the frequency plot, which is significantly different, is shown explicitly in Fig. \ref{fig_cs2_freq_const}. Two GFMIs operate at different frequencies until they synchronize with the TG, one at 12:15 and the other at 12:30. Furthermore, this method maintains both the RoCoF and $f^{nad}$ within secure limits.


\subsubsection{Comparison between proposed and benchmark methods.}  
\begin{figure}[t]
    \centering
    \includegraphics[width=\linewidth]{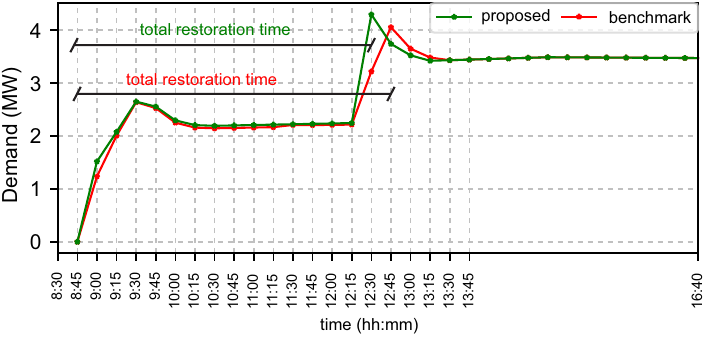}
    \vspace{-9mm}
    \caption{Comparison of restoration metrics for 2 GFMI case.}\label{fig_rest_comp}
\end{figure}

Depending on the severity of the outage, the TG may have varying restoration times. Therefore, we evaluated the performance of the proposed and benchmark methods for different restoration instances of the TG. The restoration performance for a scenario where the TG is restored at 12:00 is shown in Figure \ref{fig_rest_comp}. It is observed that optimizing both the ESW and SSWs not only rapidly restores the load but also enhances customer-hours served. The comparison for other restoration scenarios of the TG is provided in Table \ref{tab_comp}. It is evident that the proposed method has a slight edge over the benchmark method in terms of restoration metrics for longer-duration outages. Next, we will examine another case study involving four GFMIs to highlight the significance of the proposed method.

\begin{table}[t]
\setlength{\tabcolsep}{1pt}
  \centering
  \caption{Proposed vs. benchmark methods: 2-GFMIs case}
    \begin{small}
    \begin{tabular}{ccccccc}
    \toprule
    \multirow{2}[2]{*}{\parbox{1cm}{\centering \textbf{TG restored at}}} & \multicolumn{2}{c}{{\parbox{2.2cm}{\textbf{Customer hours restored (MWh)}}}} & {\multirow{2}[2]{*}{\parbox{1cm}{\textbf{\% improved}}}} & \multicolumn{2}{c}{{\parbox{2cm}{\textbf{DS restoration time (min)}}}} & \multirow{2}[2]{*}{\parbox{0.8cm}{\textbf{impr- oved (mins)}}} \\
              \cline{2-3} \cline{5-6}
          & \multicolumn{2}{c}{} &       & \multicolumn{2}{c}{} &  \\
          & \multicolumn{1}{c}{\text{proposed}} & \multicolumn{1}{c}{\text{benchmark}} &       & \multicolumn{1}{c}{\text{proposed}} & \multicolumn{1}{c}{\text{benchmark}} &  \\
    \midrule
    10:00 & 26.99 & 26.93 & 0.22  & 105   & 105   & 0 \\
    11:00 & 26.25 & 25.95 & 1.16  & 165   & 180   & 15 \\
    12:00 & 23.67 & 23.33 & 1.46  & 225   & 240   & 15 \\
    13:00 & 21.11 & 20.74 & 1.78  & 285   & 300   & 15 \\
    14:00 & 18.05 & 17.61 & 2.5   & 345   & 360   & 15 \\
    \bottomrule
    \end{tabular}%
   \label{tab_comp}%
   \end{small}
\end{table}%

\subsection{Blackstart and load restoration with four GFMIs}
For this case study, we connected an additional two GFMIs at buses 29 and 152 in the IEEE-123-bus test system, as shown in Fig. \ref{fig_IEEE_123bus}. The allocated power and energy capacities are detailed in Table \ref{tab_BES_ratings}.  With four GFMIs, four microgrids will be formed, necessitating more SSWs. The new allocation of SSWs is also shown in Table \ref{tab_BES_ratings}. We examined the performance of the proposed and benchmark methods for the modified test system, keeping the loads and PV generation profile the same as in the previous case study. The summary of the simulation results are shown in Table \ref{tab_comp_4gfmi}. 
It is observed that there is a significant improvement in the customer hours restoration metric for longer outages (e.g., 11.46\% for TG coming online at 14:00), while the restoration time is reduced by 45 minutes.
\begin{table}[t]
\setlength{\tabcolsep}{1pt}
  \centering
  \caption{Proposed vs. benchmark methods: 4-GFMIs case}
    \begin{small}
    \begin{tabular}{ccccccc}
    \toprule
    \multirow{2}[2]{*}{\parbox{1cm}{\centering \textbf{TG restored at}}} & \multicolumn{2}{c}{{\parbox{2.2cm}{\textbf{Customer hours restored (MWh)}}}} & {\multirow{2}[2]{*}{\parbox{1cm}{\textbf{\% improved}}}} & \multicolumn{2}{c}{{\parbox{2cm}{\textbf{DS restoration time (min)}}}} & \multirow{2}[2]{*}{\parbox{0.8cm}{\textbf{impr- oved (mins)}}} \\
              \cline{2-3} \cline{5-6}
          & \multicolumn{2}{c}{} &       & \multicolumn{2}{c}{} &  \\
     & \multicolumn{1}{c}{\text{proposed}} & \multicolumn{1}{c}{\text{benchmark}} &       & \multicolumn{1}{c}{\text{proposed}} & \multicolumn{1}{c}{\text{benchmark}} &  \\
    \midrule
    10:00 & 27.86 & 26.57 & 4.86  & 105   & 135   & 30 \\
    11:00 & 26.59 & 25.28 & 5.18  & 165   & 195   & 30 \\
    12:00 & 23.96 & 22.64 & 5.83  & 225   & 270   & 45 \\
    13:00 & 21.35 & 19.54 & 9.26  & 285   & 330   & 45 \\
    14:00 & 18.29 & 16.41 & 11.46   & 345   & 390   & 45\\
    \bottomrule
    \end{tabular}%
   \label{tab_comp_4gfmi}%
   \end{small}
\end{table}%

\section{Conclusion}

This paper proposes a comprehensive bottom-up framework for blackstart and load restoration planning, tailored for BES-based GFMIs and renewable GFLIs. The framework initiates blackstart using multiple GFMIs, sequentially expands microgrid boundaries to pick up cold loads, establishes cranking paths to GFLIs, synchronizes microgrids to form larger islands, and ultimately synchronizes with the TG to complete the restoration process. Integration of these steps with the linearized power flow model of the DS forms the core contribution of this study. Furthermore, we enhance GFMI models by incorporating control features from virtual synchronous generators to establish and validate quasi-steady-state and dynamic frequency responses, including RoCoF and frequency nadir. The proposed framework is validated and benchmarked on the IEEE-123-bus test system, considering configurations with two and four GFMIs across various scenarios of TG recovery times. Our findings indicate that optimizing synchronization decisions facilitates faster restoration and enhances customer hours restoration during prolonged outages, particularly with increased numbers of GFMIs.

\ifCLASSOPTIONcaptionsoff
  \newpage
\fi

\bibliographystyle{IEEEtran}

\bibliography{IEEEabrv, references.bib}

\begin{thebibliography}{10}
\providecommand{\url}[1]{#1}
\csname url@samestyle\endcsname
\providecommand{\newblock}{\relax}
\providecommand{\bibinfo}[2]{#2}
\providecommand{\BIBentrySTDinterwordspacing}{\spaceskip=0pt\relax}
\providecommand{\BIBentryALTinterwordstretchfactor}{4}
\providecommand{\BIBentryALTinterwordspacing}{\spaceskip=\fontdimen2\font plus
\BIBentryALTinterwordstretchfactor\fontdimen3\font minus
  \fontdimen4\font\relax}
\providecommand{\BIBforeignlanguage}[2]{{%
\expandafter\ifx\csname l@#1\endcsname\relax
\typeout{** WARNING: IEEEtran.bst: No hyphenation pattern has been}%
\typeout{** loaded for the language `#1'. Using the pattern for}%
\typeout{** the default language instead.}%
\else
\language=\csname l@#1\endcsname
\fi
#2}}
\providecommand{\BIBdecl}{\relax}
\BIBdecl

\bibitem{APPA}
{American Public Power Association}, ``Restoration best practices guidebook,''
  \url{https://www.publicpower.org/resource/restoration-best-practices-guidebook#:~:text=The%20Restoration%20Best%20Practices%20Guidebook,restoration%20operations%20and%20emergency%20management.},
  {(Accessed: Jun. 19, 2024)}.

\bibitem{Li2014DistributionSearch}
J.~Li, X.-Y. Ma, C.-C. Liu, and K.~P. Schneider, ``Distribution system
  restoration with microgrids using spanning tree search,'' \emph{IEEE
  Transactions on Power Systems}, vol.~29, no.~6, pp. 3021--3029, 2014.

\bibitem{Sharma2015AIslanding}
A.~Sharma, D.~Srinivasan, and A.~Trivedi, ``A decentralized multiagent system
  approach for service restoration using dg islanding,'' \emph{IEEE
  Transactions on Smart Grid}, vol.~6, no.~6, pp. 2784--2793, 2015.

\bibitem{Wang2019CoordinatingSystems}
Y.~Wang, Y.~Xu, J.~He, C.~C. Liu, K.~P. Schneider, M.~Hong, and D.~T. Ton,
  ``{Coordinating multiple sources for service restoration to enhance
  resilience of distribution systems},'' \emph{IEEE Transactions on Smart
  Grid}, vol.~10, no.~5, pp. 5781--5793, 2019.

\bibitem{Chen2018SequentialMicrogrids}
B.~Chen, C.~Chen, J.~Wang, and K.~L. Butler-Purry, ``{Sequential Service
  Restoration for Unbalanced Distribution Systems and Microgrids},'' \emph{IEEE
  Transactions on Power Systems}, vol.~33, no.~2, pp. 1507--1520, 2018.

\bibitem{Ding2022ANetworks}
T.~Ding, Z.~Wang, M.~Qu, Z.~Wang, and M.~Shahidehpour, ``{A Sequential
  Black-Start Restoration Model for Resilient Active Distribution Networks},''
  \emph{IEEE Transactions on Power Systems}, vol.~37, no.~4, pp. 3133--3136,
  2022.

\bibitem{Liu2021CollaborativeDERS}
W.~Liu and F.~Ding, ``{Collaborative Distribution System Restoration Planning
  and Real-Time Dispatch Considering Behind-the-Meter DERS},'' \emph{IEEE
  Transactions on Power Systems}, vol.~36, no.~4, pp. 3629--3644, 2021.

\bibitem{Arif2022SwitchingRestoration}
A.~Arif, B.~Cui, and Z.~Wang, ``{Switching Device-Cognizant Sequential
  Distribution System Restoration},'' \emph{IEEE Transactions on Power
  Systems}, vol.~37, no.~1, pp. 317--329, 1 2022.

\bibitem{Chen2019TowardRestoration}
B.~Chen, Z.~Ye, C.~Chen, and J.~Wang, ``{Toward a MILP Modeling Framework for
  Distribution System Restoration},'' \emph{IEEE Transactions on Power
  Systems}, vol.~34, no.~3, pp. 1749--1760, 2019.

\bibitem{Gao2022DecentralizedResources}
X.~Gao, R.~R. Nejad, and W.~Sun, ``{Decentralized Distribution System
  Restoration with Grid-Forming/Following Inverter-Based Resources},''
  \emph{IEEE Power and Energy Society General Meeting}, vol. 2022-July, pp.
  1--5, 2022.

\bibitem{Che2019AdaptiveConditions}
L.~Che and M.~Shahidehpour, ``{Adaptive Formation of Microgrids With Mobile
  Emergency Resources for Critical Service Restoration in Extreme
  Conditions},'' \emph{IEEE Transactions on Power Systems}, vol.~34, no.~1, pp.
  742--753, 2019.

\bibitem{Zhang2021AConstraints}
Q.~Zhang, Z.~Ma, Y.~Zhu, and Z.~Wang, ``{A Two-Level Simulation-Assisted
  Sequential Distribution System Restoration Model with Frequency Dynamics
  Constraints},'' \emph{IEEE Transactions on Smart Grid}, vol.~12, no.~5, pp.
  3835--3846, 2021.

\bibitem{Du2022Black-StartMicrogrids}
Y.~Du, H.~Tu, X.~Lu, J.~Wang, and S.~Lukic, ``{Black-Start and Service
  Restoration in Resilient Distribution Systems With Dynamic Microgrids},''
  \emph{IEEE Journal of Emerging and Selected Topics in Power Electronics},
  vol.~10, no.~4, pp. 3975--3986, 2022.

\bibitem{Liu2023UtilizingRestoration}
F.~Liu, C.~Chen, C.~Lin, G.~Li, H.~Xie, and Z.~Bie, ``{Utilizing Aggregated
  Distributed Renewable Energy Sources With Control Coordination for Resilient
  Distribution System Restoration},'' \emph{IEEE Transactions on Sustainable
  Energy}, vol.~14, no.~2, pp. 1043--1056, 2023.

\bibitem{Nerc_reliability}
{North American Electric Reliability Corporation (NERC)}, ``Standard prc-024-2
  — generator frequency and voltage protective relay settings,''
  \url{https://www.nerc.com/pa/Stand/Reliability%20Standards/PRC-024-2.pdf},
  {(Accessed: Mar. 23, 2024)}.

\bibitem{Ambia2021}
M.~N. Ambia, K.~Meng, W.~Xiao, A.~Al-Durra, and Z.~Y. Dong, ``{Interactive Grid
  Synchronization-Based Virtual Synchronous Generator Control Scheme on Weak
  Grid Integration},'' \emph{IEEE Transactions on Smart Grid}, vol.~13, no.~5,
  pp. 4057--4071, 2021.

\bibitem{Liu2024}
L.~Liu, Z.~Hu, Y.~Wen, and Y.~Ma, ``{Modeling of Frequency Security Constraints
  and Quantification of Frequency Control Reserve Capacities for Unit
  Commitment},'' \emph{IEEE Transactions on Power Systems}, vol.~39, no.~1, pp.
  2080--2092, 2024.

\bibitem{Kim2022VoltageInverters}
T.~Kim, N.~G. Barry, W.~Kim, S.~Santoso, W.~Wang, R.~C. Dugan, and
  D.~Ramasubramanian, ``{Voltage Balancing Capability of Grid-Forming
  Inverters},'' \emph{IEEE Open Access Journal of Power and Energy}, vol.~9,
  no. December, pp. 479--488, 2022.

\bibitem{Rathnayake2022}
D.~B. Rathnayake, R.~Razzaghi, and B.~Bahrani, ``{Generalized Virtual
  Synchronous Generator Control Design for Renewable Power Systems},''
  \emph{IEEE Transactions on Sustainable Energy}, vol.~13, no.~2, pp.
  1021--1036, 2022.

\bibitem{Lee_converters}
C.-T. Lee, R.-P. Jiang, and P.-T. Cheng, ``A grid synchronization method for
  droop controlled distributed energy resources converters,'' in \emph{2011
  IEEE Energy Conversion Congress and Exposition}, 2011, pp. 743--749.

\bibitem{CASTRO2015300}
\BIBentryALTinterwordspacing
P.~M. Castro, ``Tightening piecewise mccormick relaxations for bilinear
  problems,'' \emph{Computers \& Chemical Engineering}, vol.~72, pp. 300--311,
  2015, a Tribute to Ignacio E. Grossmann. [Online]. Available:
  \url{https://www.sciencedirect.com/science/article/pii/S0098135414001069}
\BIBentrySTDinterwordspacing

\bibitem{gazijahani2022parallel}
F.~S. Gazijahani, J.~Salehi, and M.~Shafie-Khah, ``A parallel fast-track
  service restoration strategy relying on sectionalized interdependent
  power-gas distribution systems,'' \emph{IEEE Transactions on Industrial
  Informatics}, vol.~19, no.~3, pp. 2273--2283, 2022.

\bibitem{Cheng2022}
R.~Cheng, Z.~Wang, Y.~Guo, and Q.~Zhang, ``{Online Voltage Control for
  Unbalanced Distribution Networks Using Projected Newton Method},'' \emph{IEEE
  Transactions on Power Systems}, vol.~37, no.~6, pp. 4747--4760, 2022.

\bibitem{wang2020radiality}
Y.~Wang, Y.~Xu, J.~Li, J.~He, and X.~Wang, ``On the radiality constraints for
  distribution system restoration and reconfiguration problems,'' \emph{IEEE
  Transactions on Power Systems}, vol.~35, no.~4, pp. 3294--3296, 2020.

\end{thebibliography}

\appendix

\subsection{Terminal voltage of GFMI considering voltage droop} 
\label{app_gfmi_vol}
Terminal voltage of GFMI deviates from the reference value ($V^*$) with reactive power output ($Q$) and voltage droop gain ($K^v$), which is define as:
\begin{align}
\begin{small}
    |V|= V^* - K^v Q
\end{small}
\end{align}
To comply with the branch-flow model, we define the terminal voltage of GFMI in terms of $v=|V|^2$, as:
\begin{align}
\begin{small}
    v = (V^*-K^vQ)^2 = (V^*)^2 - 2V^*K^vQ+ (K^vQ)^2 \label{eqn_gmfi_vq}.
\end{small}
\end{align}
In the square form, the deviation from $(V^*)^2$ is non-linear with Q. Hence, we simplify (\ref{eqn_gmfi_vq}) by defining a new variable $\Delta v^{cc}$, which represents the last two terms in (\ref{eqn_gmfi_vq}). With this, the terminal voltage is written as:
\begin{align}
\begin{small}
     v =(V^*)^2 + \Delta v^{cc}
\end{small}
\end{align}

\subsection{Calculation of $\gamma$ to estimate $f^{nad}$ for a VSG} \label{app_gamma}
The calculation of $\gamma$ is based on VSG control in Fig. \ref{fig_vsm_ctrl}a and work presented for the synchronous generator in \cite{Liu2024}, and is expressed as:
\begin{small}
\begin{align}
\gamma_i &= \alpha_i \sqrt{1-\xi_i^2} e^{-\xi_i\omega^o_i t_i^{nad}}\\
\text{where,}\; \alpha_i &= \sqrt{\frac{T_i K^f_i}{2H_i}}, \;
    \omega^o_i = \frac{D_i+K_i^f}{2H_i T_i}, \;
    \xi_i = \frac{2H_i + D_iT_i}{2(D_i+K^f_i)}\omega_i^o\nonumber\\ t_i^{nad}&=\frac{1}{\omega^r_i}tan^{-1}\bigg(\frac{\omega_i^rT_i}{\xi_i\omega^o_iT_i-1}\bigg),\; \omega^r_i=\omega^o_i\sqrt{1-\xi^2}\nonumber
\end{align}
\end{small}
Although calculation of $\gamma_i$ is highly non-linear, it is a constant parameter and do not increase the complexity of the proposed blackstart optimization problem.

\begin{table}[t]
  \centering
  \caption{VSG-based GFMI parameters }
  \begin{small}
    \begin{tabular}{|c|c|}
    \hline
    Parameters & values in pu\\
    \hline
      ($H, D, K^f, \gamma$) & (4,1, 89, 0.093)  \\
    \hline
    \end{tabular}%
  \label{tab_GFMI_parameters}%
  \end{small}
\end{table}%
\end{document}